\documentclass[11pt]{amsart}
\usepackage{algorithm,algorithmic}
\usepackage{amssymb,amsmath}
\usepackage{graphics}

\newcommand{\bru}{B^{\mathcal{D},n+1}}
\newcommand{\gbru}{G^{\mathcal{D},n}}
\newtheorem{thm}{Theorem}
\newtheorem{lem}{Lemma}
\begin{document}
\title{Minimal Eulerian trail in a labeled digraph}%

\author{Mart\'{\i}n Matamala}
\email{mmatamal@dim.uchile.cl}
\author{Eduardo Moreno}
\email{emoreno@dim.uchile.cl}

\thanks{Partially supported by ECOS C00E03 (French-Chilean Cooperation),
Programa Iniciativa Cient\'{\i}fica Milenio P01-005, and CONICYT
Ph.D. Fellowship.}

\address{Departamento de Ingenier\'{\i}a Matem\'atica,
Facultad de Ciencias F\'{\i}\-si\-cas y Matem\'aticas,
Universidad de Chile, Centro de Modelamiento Ma\-te\-m\'a\-ti\-co, 
UMR 2071, UCHILE-CNRS,
Casilla 170-3, Correo 3, Santiago, Chile.
}

\begin{abstract}
Let $G$ be an Eulerian directed graph with an arc-labeling such that arcs 
going out from the same vertex have different labels. In this
work, we present an algorithm to construct the Eulerian trail
starting at an arbitrary vertex $v$ of minimum lexicographical label among
labels of all Eulerian trails starting at this vertex.

We also show an application of this algorithm to construct the
minimal de Bruijn sequence of a language.
\end{abstract}


\maketitle


\section{Introduction}
Eulerian graphs were an important concept in the beginning of the
graph theory. The ``K\"onigsberg bridge problem" and its solution
given by Euler in 1736 are considered the first paper of what is
nowadays called \textit{graph theory}.

In this work, we consider graphs with an arc-labeling with the
following property: Arcs going out from the same vertex have
different labels. These graphs are commonly utilized in the automata
theory: a labeled digraph represents deterministic automata where
vertices are the states of the automata, and arcs represent the
transitions from one state to another, depending on the label of the
arc. Eulerian trails over these graphs are related with
synchronization of automata (see \cite{Kari-sync-automata}).

Eulerian graphs with this kind of labeling are also  used in the
study of DNA. By DNA sequencing we can obtain fragments of DNA which
need to be assembled in the correct way. To solve this problem, we
can simply construct a \textit{DNA graphs} (see
\cite{MR2000j:92015}) and find an Eulerian trail over this graph.
This strategy is already implemented and it is now one of the more
promising algorithms for DNA sequencing (see
\cite{Pevzner89,Pevzner01}).

To find the Eulerian trail of minimal label is also interesting
with respect to the problem of finding of optimal encoding for DRAM address bus. In this
model, an address space of size $2^{2n}$ is represented as labels of
edges in a complete graph with $2^n$ vertices. An Eulerian trail
over this graph produces an optimal multiplexed code  (see
\cite{cheng}). If we want to give priority to some address in
particular, the Eulerian cycle of minimal label give us this code.

 Another interesting application
of these graphs is to find \textit{de Bruijn sequences} of a
language. De Bruijn sequences are also known as ``shift register
sequences'' and were originally studied in
\cite{deBruijn46:a_combinatorial} by N. G. De Bruijn for the binary
alphabet. These sequences have many different applications, such as
memory wheels in computers and other technological device, network
models, DNA algorithms, pseudo-random number  generation and modern
public-key cryptographic schemes, to mention a few (see
\cite{MR22:7945,MR98e:05102,MR93m:05018}). More details about this
application are discussed in Section 3.

By the BEST theorem (see \cite{tutte84:_graph_theor}), we can
compute the number of Eulerian trails in a graph. This number is
usually exponential in the number of vertices of the graph (at least
$((\gamma-1)!)^{|V|}$ where $V$ is the set of vertices and $\gamma$
is the minimum degree of vertices in $V$ ). Therefore, finding the
Eulerian trail of lexicographically minimum label can be costly.

In this work, we give an algorithm to construct the Eulerian trail
of minimum label starting at a given vertex. The complexity of the
algorithm is linear in the number of arcs of the graph.  In Section
2 we give some definitions to understand the problem and we prove
the main theorem. Finally, in Section 3 we give an application of
this algorithm to construct the minimal de Bruijn sequence of a
language.

\section{Main Theorem}
Let $G$ be a digraph  and let $l:A(G)\to N$ be a labeling of the
arcs of $G$ over an alphabet $N$ such that arcs going out from the
same vertex have different labels.

A \textit{trail} is an alternating sequence $W=v_1 a_1 v_2 a_2
\ldots v_{k-1} a_{k-1} v_k$ of vertices $v_i$ and arcs $a_j$ such
that the tail of $a_i$ is $v_i$ and the head of $a_i$ is $v_{i+1}$
for every $i=1,2, \ldots, k-1$ and all arcs are distinct. If
$v_1=v_k$ then $W$ is a closed trail. A closed trail is an Eulerian
trail if the arcs of $W$ are all the arcs  of $G$. An Eulerian graph is
a graph with an Eulerian trail. The label of $W$ is the word
$l(a_1)\ldots l(a_{k-1})$.

Given a strongly connected Eulerian digraph and a vertex $r$, we show how to find the Eulerian trail starting in $r$ with the minimal lexicographical label. Remark that is important to fix a starting vertex $r$ so at to define an order in which vertices are visited, which allow us to define a lexicographical order among Eulerian trails.

Let $U$ be a subset of vertices in $G$. A \textit{cut} is the set of arcs with one end
in $U$ and the other in $ V\setminus U$, and is denoted by $\delta_G(U)$.
 A vertex $v$ is {\it exhausted} by a trail $W$ if
$\delta_{G\setminus A(W)}(v)=\emptyset$. The set of vertices
exhausted by $W$ is denoted by $S(W)$. 

\begin{lem}\label{exhausted}
Let $U,B$ be subsets of vertices and let $T$ be the trail starting in $r$
of minimum label exhausting $U$. 
If $U\subseteq B \subseteq S(T)$, then $T$
is the trail of minimum label exhausting $B$.
\end{lem}
\begin{proof}
Let  $T'$ be a trail starting in $r$ exhausting $B$ with a smaller label than $T$.
Since $U\subseteq B$ then $T'$ exhausts $U$. Hence, the label of $T$
is not minimal. 
\end{proof}

A trail $W$ can visit a vertex $v$ many times. We decompose a
trail $W$ in the sub-trails $W v$ and $v W$,  where $W v$ is the
sub-trail of $W$ finishing in the \emph{last} visit of $v$, and $v
W$ is the sub-trail of $W$ starting from the \emph{last} visit of
$v$. We denote $\overset{\circ}{v}W$ the trail $v W$ without the
first vertex $v$ but containing the first arc of $v W$.

\begin{lem}\label{corte}
Let $T$ be a closed trail starting in $r$ such that 
$r\in S(T)$. Let $v$ be the last vertex not belonging to $S(T)$ visited by $T$. If $w$ is its next vertex in $T$ then
\[ \delta_{G\setminus A(Tv)}(\overset{\circ}{v}T)=\{vw\} \]
\end{lem}
\begin{proof}
Let $xy$ be an arc of $\delta_{G}(\overset{\circ}{v}T)$. Since all
vertices of $\overset{\circ}{v}T$ are exhausted by $T$, $xy\in
A(T)$.  Hence either $xy\in A(Tv)$ or $xy\in A(vT)$. Therefore
$xy\in \delta_{G\setminus A(Tv)}(\overset{\circ}{v}T)$ if and only
if $xy=vw$. 
\end{proof}

We define the following strategy to construct a trail: Starting at a
given vertex $v$, follow the unvisited arc (if exists) of minimal
label. This strategy finishes with a closed trail, and this trail
exhausts the vertex $v$. A trail constructed by this strategy
is called an \textit{alphabetic trail} starting at $v$ and is 
denoted by $W(G,v)$. By definition, an alphabetic trail starting
at $v$ is the trail of minimal label among all trails starting at
$v$ and exhausting $v$.

Let $v$ be a vertex and let $T$ be the closed trail of minimal label
exhausting all vertices in $\overset{\circ}{v}T$. We find the
trail of minimal label exhausting all vertices in $vT$. If $v\in
S(T)$ then by Lemma \ref{exhausted} the trail $T$ is the solution to
this problem. If $v \notin S(T)$ then the next lemma give us the
solution: we need to split $T$ and insert the alphabetic trail over
$G\setminus A(T)$ starting at $v$. Repeating this process we 
finish with the Eulerian trail of minimal label.

\begin{lem}\label{main}
Let $T$ be a closed trail starting and exhausting $r$ such that 
if $v$ is the
last vertex in $V\setminus S(T)$ visited by $T$ then $T$ is the
closed trail of minimum label exhausting $\overset{\circ}{v}T$.

Let $Z$ be the closed trail of minimum label in exhausting $v T$ and
let $W=W(G\setminus A(T),v)$. Then $Z=(Tv)W(vT)$.
\end{lem}
\begin{proof}
By supposition, $T$ is the closed trail of minimum label exhausting
$\overset{\circ}{v}T$ and $\overset{\circ}{v}T\subset S(Z)$, hence
$l(Z)\geq l(T)$. In particular, $l(Z)\geq
l(Tv)$. Also $Z$
 and $(Tv)W(vT)$ exhausts $vT$. Hence $l(Z)\leq
l((Tv)W(vT))$, concluding that $Z=(Tv)Z'$.

By Lemma \ref{corte} the only way to visit vertices in
$\overset{\circ}{v}T$ is using the arc $vw$, and
$\overset{\circ}{v}T$ is the trail of minimum label exhausting
$V(\overset{\circ}{v}T)$ in $G\setminus(A(Tv))$. Since $Z$ is a
closed trail of minimum label, $Z=(Tv)Z''(vT)$.

Finally, $Z''$ is a closed trail of minimum label in $G\setminus
A(T)$ exhausting $v$, therefore $Z''=W$. 
\end{proof}

%
\begin{algorithm}
\caption{Compute the minimal Eulerian trail starting in $r$}
\label{alg}
\begin{algorithmic}
\STATE $T \leftarrow \emptyset$ \STATE $v \leftarrow
$\textsc{NoEx}($T$) \ \ \COMMENT{$v=r$} \WHILE{$v \neq$
\textbf{NULL}}
\STATE $W \leftarrow W(G\setminus A(T),v)$.
over $G\setminus A(T)$. \STATE $T \leftarrow \ (T v) W (v T)$
\STATE $v \leftarrow $ \textsc{NoEx}($T$).
\ENDWHILE
\end{algorithmic}
Where \textsc{NoEx}($T$) returns the last non-exhausted vertex
visited by $T$ or  \textbf{NULL} if this vertex does not exist.
\end{algorithm}

\begin{thm}
Algorithm \ref{alg} finishes with an Eulerian trail starting in $r$
and its label is the minimal one among all Eulerian trails starting
in $r$.
\end{thm}

\begin{proof}
At each repetition of the ``while'', the trail $T$ exhausts at least
one vertex non-exhausted in the previous step, so the algorithm finishes
 in a finite number of steps.

We define  inductively $G^i=G\setminus A(T^{i-1})$,
$v^i=$\textsc{NoEx}$(T_{i})$, $W^i=W(G^i,v^i)$ and
$T^i=(T^{i-1}v^{i-1}) W^i (v^{i-1} T^{i-1})$, with $T_0=\emptyset$.

We prove by induction that $T^i$ is the closed trail of minimal
label exhausting $\overset{\circ}{v}^i T^i$. For $i=1$,
$T^1=W(G,r)$ is by definition the closed trail of minimal label
exhausting $r$, and by Lemma \ref{exhausted} it is the trail of
minimal label exhausting $\overset{\circ}{v}^1T^1$. Let $T^{i-1}$ be
the closed trail of minimum label exhausting
$\overset{\circ}{v}^{i-1}T^{i-1}$.  Applying Lemma \ref{main} to
$T^{i-1}$, we conclude that $T^i$ is the closed trail of minimal
label exhausting $v^{i-1}T^i$ and by Lemma \ref{exhausted} it is the
minimal closed trail exhausting $\overset{\circ}{v}^i T^i$.

Therefore the algorithm finishes with a closed trail $T$
exhausting all its vertices $V(T)$, but $G$ has only one strongly
connected component, thus $A(T)=A(G)$. We conclude that $T$ is an
Eulerian trail of minimal label. \hfill\qed\end{proof}

We can use the following structure to represent the graph, a list of size $|V|$ representing vertices where each element $v$ in the list has a stack with the head of each arc starting at $v$ in order. Knowing this structure of a graph, the algorithm can easily construct the trails $W(\cdot,\cdot)$, removing the visited nodes from the stack and keeping track of exhausted vertices. Since this algorithm visits each arc at most twice, it can be implemented in $\mathcal{O}(|A(G)|)$, which is best possible.

\begin{figure}[ht]
\begin{center}
\includegraphics{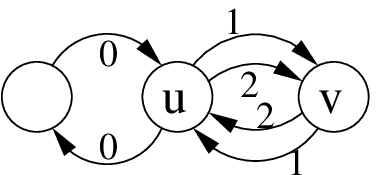}
\caption{}\label{F:exa2}
\end{center}
\end{figure}

Remark that while the initial  vertex $r$ can be arbitrarily chosen, different initial vertices can produce different trails, even if we consider the label as a circular string. For example, in the graph of Figure \ref{F:exa2}, the minimal de Bruijn sequence starting at $u$ is $001122$ but starting at $v$ is $100122$. 

\section{An application: minimal de Bruijn sequence}
Given a set $\mathcal{D}$ of words of length $n$, a de Bruijn
sequence of span $n$  is a periodic sequence $B$ such that every  word
in $\mathcal{D}$ (and no other $n$-tuple) appears exactly once in $B$.
Historically, de Bruijn sequence was studied in an arbitrary
alphabet considering the language of all the $n$-tuples.
In \cite{moreno05:dB_general_languages} the concept
of de Bruijn sequences was generalized to restricted languages with
a finite set of forbidden substrings and it was proved the existence
of these sequences and presented an algorithm to generate one of
them. Nevertheless, it remained to find the minimal de Bruijn
sequence in this general case.

In \cite{moreno04:minimaldebruijn} was studied
some particular cases where it is possible to obtain efficiently the
minimal de Bruijn sequence. Using our previous algorithm we can
solve this problem efficiently in all cases.

A word $p$ is said to be a {\it factor} of a word $w$ if there exist
words $u,v\in N^*$ such that $w=u p v$. If $u$ is the empty word
(denoted by $\varepsilon$), then $p$ is called a {\it prefix} of
$w$, and if $v$ is empty then is called a {\it suffix} of $w$.

Let $\mathcal{D}$ be a set of words of length $n+1$. We call
this set a \textit{dictionary}.  A \textit{de Bruijn
  sequence of span $n+1$} for $\mathcal{D}$  is a (circular) word $\bru$ of
length $|\mathcal{D}|$ such that all the words in $\mathcal{D}$ are
factors of $\bru$. In other words,
\[\{ (\bru)_i\ldots (\bru)_{i+n \mod (n+1)} | i=0\ldots
|\mathcal{D}|\}=\mathcal{D}\]

De Bruijn sequences are closely related to de Bruijn graphs.
 The \textit{de Bruijn graph of span $n$}, denoted by $\gbru$, is the
 directed graph with vertex set
\[ V(\gbru)=\left\{ u\in N^n |  u \mbox{ is a prefix or a suffix
 of a word in }\mathcal{D} \right\}\]
and arc set
\[ A(\gbru)=\left\{ (\alpha v,v\beta) | \alpha,\beta\in N,
\alpha v \beta\in \mathcal{D}\right\}\]

Note that the original definitions of de Bruijn sequences and de
Bruijn graph given in \cite{deBruijn46:a_combinatorial} are the
particular case of $\mathcal{D}=N^{n+1}$.

We label the arcs of the graph $\gbru$ using the following function $l$: if
$e=(\alpha u,u \beta)$  then $l(e)=\beta$.  This labeling has an
interesting property:
Let  $T=v_0 e_0\ldots e_m v_{m+1}$ be a trail over $\gbru$ of length $m\geq n$.
Then $T$ finishes in a vertex $u$ if and only if  $u$ is a  suffix
of $l(T)=l(e_0)\ldots l(e_m)$.
%
This property explains the relation between de Bruijn graphs and de
Bruijn sequence:
%
$\bru$  is the label of an Eulerian trail of $\gbru$.
%
%
Therefore, given a dictionary $\mathcal{D}$, the  existence of  a de
Bruijn sequence of span $n+1$ is characterized by the existence of
an Eulerian trail over  $\gbru$.

Let $D$ be a dictionary such that $\gbru$ is an Eulerian graph. Let
$z$ be the vertex of minimum label among all vertices.  Clearly, the
minimal de Bruijn sequence has $z$ as prefix. Hence, the minimal
Eulerian trail over $\gbru$ starts at an (unknown) vertex and after
$n$ steps it arrives to $z$. Therefore if we start our Algorithm
\ref{alg} in the vertex $z$ we obtain the Eulerian trail of minimal
label starting at $z$ which have label $B=B'\cdot z$. Hence $z\cdot
B'$ is the minimal de Bruijn sequence of span $n+1$ for
$\mathcal{D}$.

\bibliographystyle{elsart-num}
\bibliography{../DinSimb,../../CV/yo-ISI,../../CV/yo}

\end{document}